\documentclass[sigconf, nonacm, 10pt]{acmart}

\makeatletter
\def\@ACM@checkaffil{
    \if@ACM@instpresent\else
    \ClassWarningNoLine{\@classname}{No institution present for an affiliation}%
    \fi
    \if@ACM@citypresent\else
    \ClassWarningNoLine{\@classname}{No city present for an affiliation}%
    \fi
    \if@ACM@countrypresent\else
        \ClassWarningNoLine{\@classname}{No country present for an affiliation}%
    \fi
}
\usepackage{pdflscape}
\usepackage{algorithm}
\usepackage{algpseudocode}
\usepackage{graphicx} 
\usepackage{subcaption}
\usepackage{booktabs}
\usepackage{wrapfig}
\usepackage{colortbl}
\usepackage{enumitem}
\usepackage{wrapfig} 
\usepackage[justification=centerlast]{caption}
\usepackage{pifont}
\usepackage{stfloats} 
\usepackage{cleveref}
\usepackage{caption}
\crefformat{section}{\S#2#1#3}
\usepackage[all=normal,wordspacing]{savetrees}
\newcommand{\sysname}{\texttt{Cruise Control}}

\newcommand{\eg}{{\it e.g.}}
\newcommand{\ie}{{\it i.e.}}
\newcommand{\etal}{{\it et al.}}
\renewcommand{\paragraph}[1]{\vspace*{0.03in}\noindent\textbf{#1}}

\definecolor{Gray}{gray}{0.9}
\renewcommand\footnotetextcopyrightpermission[1]{}
\title{\sysname{}: Dynamic Model Selection for ML-Based Network Traffic Analysis}

\author{Johann Hugon}
	\affiliation{%
	\institution{École Normale Supérieure de Lyon}
}

\author{Paul Schmitt}
	\affiliation{%
	\institution{California Polytechnic State University}
}

\author{Anthony Busson}
	\affiliation{%
	\institution{Université Lyon 1}
}

\author{Francesco Bronzino}
	\affiliation{%
	\institution{École Normale Supérieure de Lyon}
}

\begin{document}

\begin{abstract}
As networks grow increasingly complex, machine learning (ML) has become
indispensable for real-time operational insights, from traffic classification to
quality monitoring and threat detection. However, these ML-based systems face
critical computational constraints in production environments, where exceeding
resource limitations can lead to packet loss and compromise performance.
Existing approaches that use static model configurations face two significant
challenges: they require operators to possess detailed knowledge of deployment
environments and associated system budgets, making configuration tedious and
error-prone; and they must be configured for worst-case scenarios, leading to
resource inefficiency during normal operations. Because of these limitations,
deploying ML models in real-world networks with tight performance constraints
remains an open challenge. 

In this paper, we present a system-driven approach to dynamically select ML
models based on their computational complexity and available system
capacity---in real-time. Our proof-of-concept implementation, \sysname{},
supports multiple parallel ML tasks with efficient resource allocation across
multiple analytical functions. Experimental results using two real-world traffic
analysis tasks show \sysname{} improves median task accuracy by 2.78\% while
reducing packet loss by a factor of four compared to statically-selected models.
In addition, ~\sysname{} offers an easy-to-configure solution for extracting
multiple tasks, while effectively combining them and selectively collecting
features.

\end{abstract}

\maketitle
\section{Introduction}\label{sec:intro}

Machine learning (ML) has rapidly become an indispensable tool for network
traffic analysis tasks, thanks to ML models' excellence at discovering complex
relationships between network traffic and critical events occurring across
different network layers. Consequently, the networking community has
increasingly developed sophisticated ML-based solutions to assist with essential
tasks such as traffic
classification~\cite{bernaille2006earlyappid,shapira2021flowpic,rezaei2019deepappid,ACDC,piet2023ggfast},
Quality of Experience (QoE)
inference~\cite{mangla2018emimic,bronzino2019inferring,sharma2023estimating,mangla2018quicqoe},
intrusion detection~\cite{khraisat2019survey,ahmad2021network,liu2019machine},
and numerous other critical network analysis
functions~\cite{singh2013mlsurvey,nguyen2008mlsurvey,boutaba2018mlsurvey}.
However, deployment of ML models in operational networks remains a considerable
challenge due to the dynamic nature of network traffic and the system
requirements imposed by the need to process large volumes of data in real-time.

ML-based traffic analysis deployments typically employ a multi-stage pipeline
that operates under strict real-time constraints~\cite{TrafficRefinery,CATO}.
This pipeline consists of three critical stages: (1) traffic collection at a
network vantage point, (2) transformation of raw packets into feature
representations suitable for ML models, and (3) model execution to generate
analytical outputs. Each stage must operate within tight processing budgets to
prevent packet loss, which can severely inhibit model performance (\ie,
accuracy). For example, this performance degradation is particularly severe for
inference tasks where discriminative features often appear in the initial packet
exchanges of a network flow~\cite{babaria2025fastflow}. Missing these early
packets can drastically reduce classification accuracy, rendering the entire
analytical pipeline ineffective.

Prior research has addressed this challenge through two main approaches. First,
by designing specialized models that minimize packet
requirements~\cite{piet2023ggfast,babaria2025fastflow}. However, models using
this approach remain vulnerable to packet loss when processing budgets are
exceeded and fail to scale to modern network speeds of tens or hundreds of
gigabit per second. The second approach focuses on developing frameworks that
help operators balance tradeoffs between model performance and system
constraints such as latency~\cite{CATO}, CPU utilization~\cite{TrafficRefinery},
and memory consumption~\cite{ACDC}. In particular, these works have targeted the
first two pipeline stages—traffic collection and traffic
transformations—demonstrating that these stages often create bottlenecks in
processing pipelines, thus requiring optimization to support downstream models.
However, these solutions are limited to targeting static deployment of a single
model during runtime.

The static model approaches face fundamental limitations: network deployment
environments exhibit significant variability in topology, traffic patterns, and
resource constraints--making it impossible to select a universally optimal
model. Given this variability, operators typically resort to deploying models
designed for worst-case traffic scenarios to avoid packet loss during peak
periods. This conservative approach introduces a clear systematic inefficiency:
during normal (\eg, non-peak) conditions, the worst-case model results in lower
accuracy than that which could be achieved by a more complex model, given system
resources and load~\cite{romero2021INFaas}. 

In this paper, we develop \sysname{}, a complementary, system-driven solution to
adaptively select the best-fit target models for one or more feature extraction
pipelines in parallel, removing the need for operators to select the best
performing models for any given environment \textit{a priori}. By leveraging
real-time insights into the current system capabilities and traffic
characteristics, \sysname{} dynamically selects the ideal target model from a
pool of candidates for each task, ensuring that the system can effectively
characterize traffic as network conditions evolve while avoiding packet loss. Of
course, shifting to such approach requires tackling new system challenges.
First, the system must be able to adapt to changing network conditions without
incurring packet loss due to the overhead of pipeline changes. Second, the
system must be able to monitor the existing processing pipeline and determine
whether it is best suited for the ongoing conditions. Finally, the system must
support running multiple heterogeneous ML tasks (each with different
accuracy-cost tradeoffs) in parallel and balance model accuracy and efficiency
across them to ensure effective traffic characterization under dynamic network
conditions.

We address these challenges through the following contributions: first,
\sysname{} minimizes the burden of model deployment by solely requiring a user
to provide a set of models with different accuracy-cost tradeoffs. Second,
\sysname{} implements a selection pipeline that enables parallel feature
collection for multiple models, minimizing overhead. Third, \sysname{} monitors
lightweight signals (\ie, packet loss) to evaluate its current processing
capacity and dynamically selects models based on the system's current
capabilities and the observed traffic. This approach allows the system to adapt
to changing conditions, ensuring that the selected model remains the best-fit under varying traffic loads.

We implement \sysname{} and evaluate its performance for two real world traffic
analysis tasks: video quality inference and traffic classification. Our results
show that \sysname{} effectively adapts to changing network conditions,
selecting the most appropriate model for the observed traffic and system
capabilities. Compared to existing approaches that aim to select an optimal
candidate model for each task based on offline information, \sysname{} reduces
packet loss by a factor of four while achieving equivalent or better median
accuracy. Further, \sysname{} supports the parallel execution of multiple
network analysis tasks, allowing for efficient resource allocation across
different analytical functions without sacrificing performance. This
parallelization ensures ease of use and reduces operational complexity, allowing
operators to perform complex tasks effortlessly, avoiding the need for multiple
processing servers.

We hope our approach to dynamic model selection will pave the way for
exploration in the path to widespread adoption of machine learning solutions for
network traffic analysis. To stimulate further research and innovation in this
direction, we release the source code for \sysname{}\footnote{The source code
will be made available upon paper acceptance.} to the community, encouraging
others to build upon our design.

\section{Background and Motivation}\label{sec:context}

Network operators rely on the ability to reason about their networks, from
understanding the traffic that traverses them to whether they are functioning
correctly. However, answering such questions is increasingly difficult due to
multiple factors, including traffic volume (\ie, relentlessly increasing network
speeds)\cite{Zhipeng2020100gbps} and opacity (\ie, widespread adoption of
encryption)\cite{Papadogiannaki2021surveyencryptednetworkanalysis,BehindTheCurtain}.
As such, conventional monitoring approaches are becoming
inadequate~\cite{Papadogiannaki2021surveyencryptednetworkanalysis,TLS1.3Survey,Zhipeng2020100gbps}
as they struggle to cope with modern Internet traffic characteristics. 

To address these challenges, network operators have turned to
ML-based solutions~\cite{boutaba2018mlsurvey,MLEncrypted,DeepLearningSurvey} for
various network monitoring tasks, from traffic classification to quality of
service inference~\cite{MLQoSQoE}. For example, while encryption makes direct
measurement of application layer performance such as video streaming quality
metrics impossible, ML models are able to accurately infer these metrics from
encrypted network flows, providing crucial insights into user
experience~\cite{Requet,mangla2018emimic,mazhar2018Qoe,EFFECTOR,bronzino2019inferring}.
Unfortunately, ML-based approaches that use static configurations fall short
as they fail to capture the inherent variability of network
traffic. In this section we discuss current approaches to network monitoring
using ML and identify challenges that operational deployments currently face,
particularly in the face of packet loss.

\subsection{ML-Based Traffic Analysis}

The typical ML-based traffic analysis pipeline follows a structured approach:
First, raw network traffic is captured and undergoes preliminary processing,
including operations such as header parsing, flow tracking, and data reassembly.
The second stage focuses on feature extraction, where statistical computations
and information encoding prepare the data for model input. Finally, the
processed features are fed into an ML model to perform the target inference
task. Traditionally, the pipeline is evaluated based on inference performance
(\eg, accuracy or F1 score). However, ML network deployments must operate in
real-time and are thus subject to systems-related constraints such as the
ability to extract packets from the network, compute features, and make
inferences at line rates. Failing to do so can lead to packet loss, which
compromises model performance~\cite{babaria2025fastflow}.

A naïve solution is to create and deploy models using
features with low computational complexity, leaving the system with processing
headroom to accommodate unexpected traffic spikes~\cite{romero2021INFaas}.
However, this approach can impose an unnecessary ceiling on model
accuracy~\cite{TrafficRefinery} (\eg, a model with higher accuracy could be
deployed when there is less traffic, and a model with lower accuracy when there
is more). 

To address this challenge, recent
works~\cite{TrafficRefinery,CATO,piet2023ggfast,babaria2025fastflow,ACDC} have
proposed approaches that holistically consider both model accuracy and system
performance. For example, CATO~\cite{CATO} highlighted that the choice of which
features to compute---potentially even more than the selection of the model
itself---can significantly affect a measurement system's ability to gather the
necessary information for effective model execution. Consequently, CATO proposed
a method to automatically generate Pareto-optimal configurations that maximize
accuracy while minimizing system resource usage for a given network environment.
However, these configurations, while statically optimal, only represent a single
network environment. Further, network traffic loads are inherently dynamic,
leading to significant load variance over time. The outcome of these factors is
that models that perform well in offline training and testing environments may
become unusable due to packet loss in real-world deployments.

\subsection{Downsides of Static Model Selection}

\begin{figure}[t] 
   \centering
   \includegraphics[width=\columnwidth]{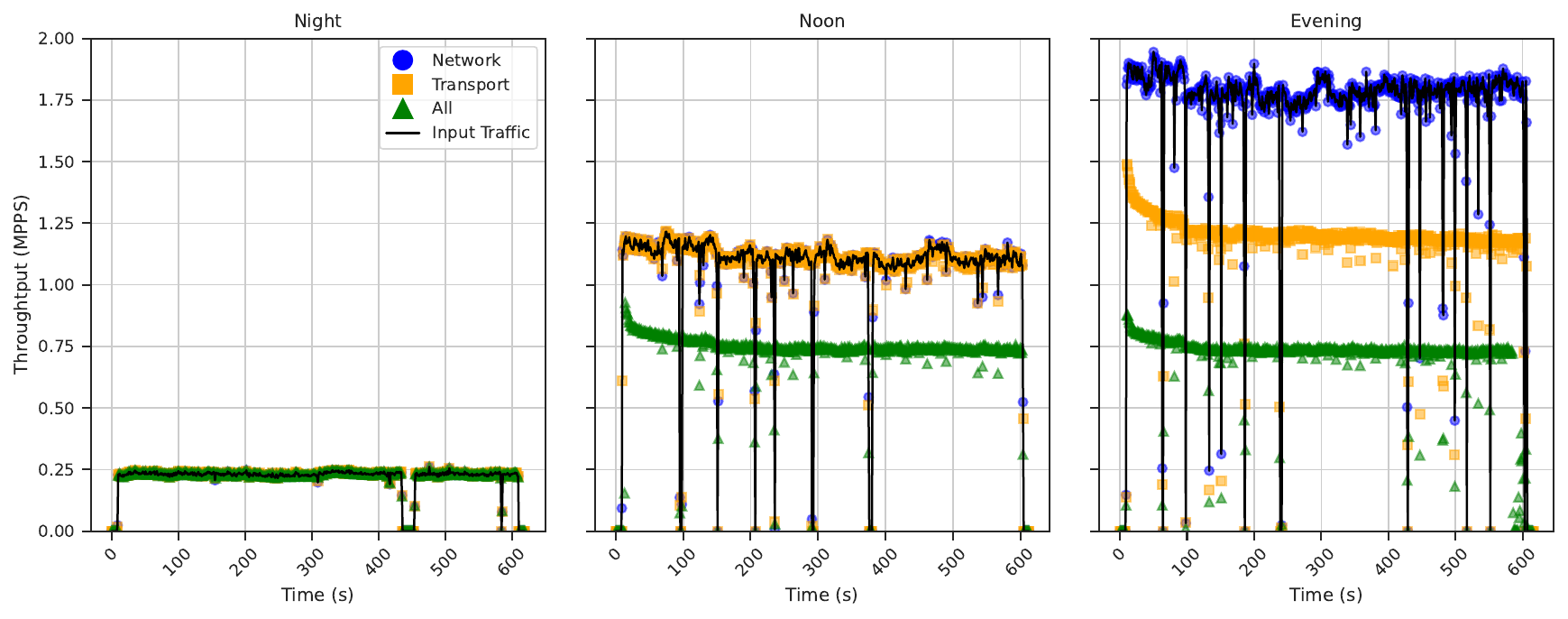}
   \caption{Comparison of the impact of three different video quality inference models across different times of day.}  
    \label{fig:different_time_combined}
\end{figure}

To illustrate the inefficiencies that can manifest using static model
deployment, we perform a small experiment using a typical ML-based traffic
analysis task: video resolution inference from encrypted
traffic~\cite{mangla2018emimic,
bronzino2019inferring,sharma2023estimating,mangla2018quicqoe}. We base our
experiment on models developed in previous work by Bronzino
\etal~\cite{bronzino2019inferring}, where the authors evaluate several feature
sets and models to infer video quality. These feature sets map to different
layers of the network stack including: \textit{Network}: basic network flow
features (\eg, throughput in/out, packet counts in/out), \textit{Transport}:
end-to-end latency and packet retransmission information (\eg, RTT,
retransmission in/out), and \textit{All}: combined features from network,
transport, and application layers (\eg, video segment sizes, time between video
segments). These different feature sets result in three models with varying
accuracy for the same task.

We evaluate the ability of an ML-based measurement system to process traffic for
these three models. As in the rest of the paper, we measure the ability of an ML
system to support a given model by analyzing whether the system can successfully
compute the features necessary for the model execution (\ie, the feature set)
without packet loss. For this experiment, we implement the different feature
sets using Retina~\cite{Retina}, a state-of-the-art feature extraction system
and the same system that CATO uses for finding Pareto-optimal configurations.
Retina enables users to efficiently compute features for subsets of parsed
flows. However, changing the feature set in Retina requires a full system
reload, a process that can take several seconds. We deploy the system on a
server equipped with a 100 Gbps network interface and evaluate its performance
using real-world traffic traces. Specifically, we use three 10-minute traces
derived from a one-hour trace collected at Equinix Chicago~\cite{caida2016}.
These traces are scaled to represent different traffic regimes throughout a
typical day, using ratios inspired by Feldmann et al.\cite{covid19}. We use
TRex\cite{TRex} to adjust the traces to reflect night ($\times 0.2$), noon
($\times 1.0$), and evening ($\times 1.6$) traffic volumes.
This is achieved by modifying the inter-packet time in the original trace,
thereby scaling the number of packets and flows. 
Additional details about the testbed setup are provided in Section~\ref{sec:eval}.

Figure~\ref{fig:different_time_combined} shows the throughput for three periods
of the day in packets processed per second by the system. Throughputs for the
feature sets are represented by blue circles, orange squares, and green
triangles respectively, while the black line represents the incoming load.
Packet loss can inferred from the difference between the input traffic and
throughput. We observe for the Night scenario, all throughputs align with the
input traffic load, indicating successful processing without packet loss for all
feature sets. Conversely, for Noon, we observe that the \textit{All} feature set
results in loss, while the other two are able to process traffic without loss.
Finally, the figure shows that in the evening the system can process traffic
without loss only for the \textit{Network} feature set, the set that is least
computationally expensive (and least accurate).

\subsection{The Accuracy Costs of Packet Loss}\label{sec:motivation}

\begin{table}[t!]
   \centering
   \scriptsize

   \begin{tabular}{llrr}
      \toprule
         $p_1$ & $1-p_2$ & \textbf{Network (MAE)} & \textbf{Transport (MAE)}\\
      \toprule
      \rowcolor{Gray} 0.0 & 0.0 & 821.8 & 560.6\\
      \midrule 
      0.005 & 0.1   & 1084.9 & 667.5\\
      \rowcolor{Gray} & 0.01 & 1334.5 & 1429.9\\
            & 0.001 & 1785.4 & 6279.5\\
      \midrule 
      \rowcolor{Gray}
      0.01  & 0.1   & 1301.4 & 714.2\\
            & 0.01  & 2322.9 & 2996.1\\
      \rowcolor{Gray}
            & 0.001 & 3596.8 & 10454.0\\
      \bottomrule 
   \end{tabular}
   
   \caption{Impact of bursty losses on video startup time inference as Median Absolute Error (MAE) in ms.}
   \label{tab:combined_loss}
\end{table}

The results of the experiment show that static model selection
introduces inherent compromises in system performance. Two solutions to this
tradeoff exist: either the operator selects a model that is guaranteed to work
under all conditions, or the operator select a model that is guaranteed to work
under most conditions, coping with possible packet loss. However, the former
approach leads to suboptimal model performance during normal operations, while
the latter approach can lead to data loss during peak periods. We demonstrate
this tradeoff using one of the two inference tasks presented by
Bronzino~\etal~\cite{bronzino2019inferring}: video startup time inference. For
this task, the first 10 seconds of features collected are used to infer the
delay between the start of the session and instant the video player actually
starts playing the video. We train three models (corresponding to the previous
feature sets) using the dataset in the paper. However, we test it both against
an unaltered test set, as well as against test sets where network traffic has
been modified via the introduction of packet loss. In particular, we introduce
bursty loss which is typical for network systems. Our bursty loss model is
based on a simple Markovian model that uses two loss states where $p_1$
represents the probability of transitioning from a no-loss state to a
state of bursty loss and $p_2$ denotes the probability of remaining in the
loss state. Results are shown in Table~\ref{tab:combined_loss}.

Two main observations can be made from the table. First, the \textit{Network}
model show, under no losses conditions, a 146.59\% higher median absolute error
compared to a model using \textit{Transport} layer features (we omit the
\textit{All} model as presented trends are just worsened), rendering it
sub-optimal. Second, bursty loss impacts faster the performance of the
\textit{Transport} model. While recent work~\cite{babaria2025fastflow} has shown
that loss can negatively impact other tasks that use single packets for
inference, this confirms that even models that use statistics over multiple
packet aggregates can be affected in the presence of bursty loss. Overall, this
experiment shows the inherent inefficiencies in model accuracy and system
performance due to static model selection and the need to avoid packet losses.
We conclude that, to achieve efficiency gains in model accuracy and system
performance, a dynamic system adaptation to network load is required.

\begin{figure*}[!ht]
    \includegraphics[width=0.75\textwidth,trim={20 30 20 7}, clip]{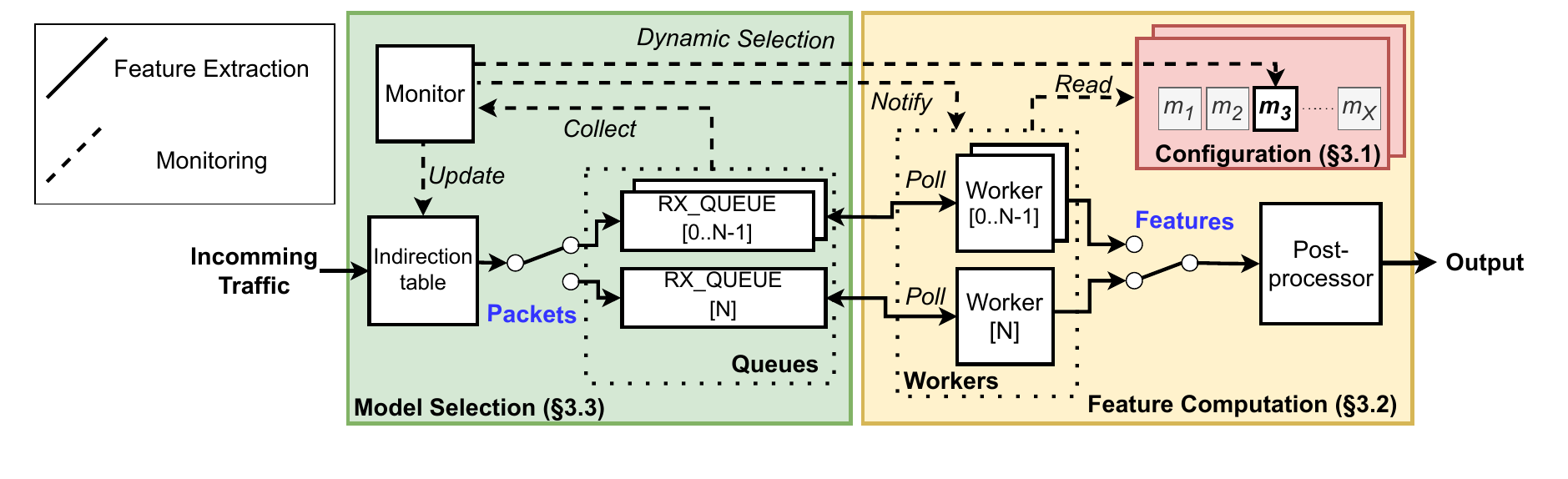}
    \caption{~\sysname{} system overview.}
    \label{fig:diagram_CC}
\end{figure*}

\section{\sysname}\label{sec:cruise_control}

In this section, we present \sysname{}, a proof-of-concept system for machine
learning-based network traffic analysis. \sysname{} is built on three key
principles: \ding{182} It simplifies deployment by requiring only a ranked list
of models per task, eliminating complex setup overheads (\cref{sec:offline}).
\ding{183} It achieves line-rate feature extraction and seamless model switching
through a coordinated architecture of workers and a post-processor core;
additionally, it minimizes multi-task processing overhead by merging common
features and eliminating redundancy (\cref{sec:feature_computation}). \ding{184}
It dynamically selects the most appropriate model in real-time using lightweight
monitoring signals, adapting to changing network conditions
(\cref{sec:model_selection}). The rest of this section details the modules
implementing these design principles, as shown in Figure~\ref{fig:diagram_CC}.

\subsection{System Configuration}\label{sec:offline}
Supporting dynamic model switching for traffic analysis requires understanding
the trade-offs between feature collection costs, model accuracy, and system
capacity under varying network conditions. \sysname{} minimizes deployment
hurdles by eliminating the need for manual fine-tuning of candidate models for
specific workloads, requiring users to solely provide a ranked list of models
and their associated feature extraction sets as input. While selecting optimal
models without prior knowledge of the deployment environment is challenging,
recent research~\cite{TrafficRefinery, CATO} has developed methods to quantify
the accuracy-cost tradeoff. We leverage CATO~\cite{CATO} to build offline input
configurations which, when coupled with \sysname{}'s dynamic model switching
capabilities, significantly reduces deployment overhead.

\begin{figure}[t!]
\begin{minipage}[c]{0.42\columnwidth}
        \centering
        \scriptsize
        \begin{tabular}{lrr}
            \toprule
            \textbf{Model \#} & \textbf{Cost} & \textbf{Acc.}\\
            \midrule
            \rowcolor{Gray}$m_9$ & 2272 & 0.935 \\
            $m_8$ & 1696 & 0.934 \\
            \rowcolor{Gray}$m_7$ & 1248 & 0.933 \\
            $m_6$ & 960 & 0.932 \\
            \rowcolor{Gray}$m_5$ & 736 & 0.931 \\
            $m_4$ & 704 & 0.926 \\
            \rowcolor{Gray}$m_3$ & 480 & 0.924 \\
            $m_2$ & 320 & 0.900 \\
            \rowcolor{Gray}$m_1$ & 256 & 0.799 \\
            \bottomrule
        \end{tabular}
        
\end{minipage}
\hfill
\begin{minipage}[c]{0.55\columnwidth}
        \centering
        \includegraphics[width=\linewidth]{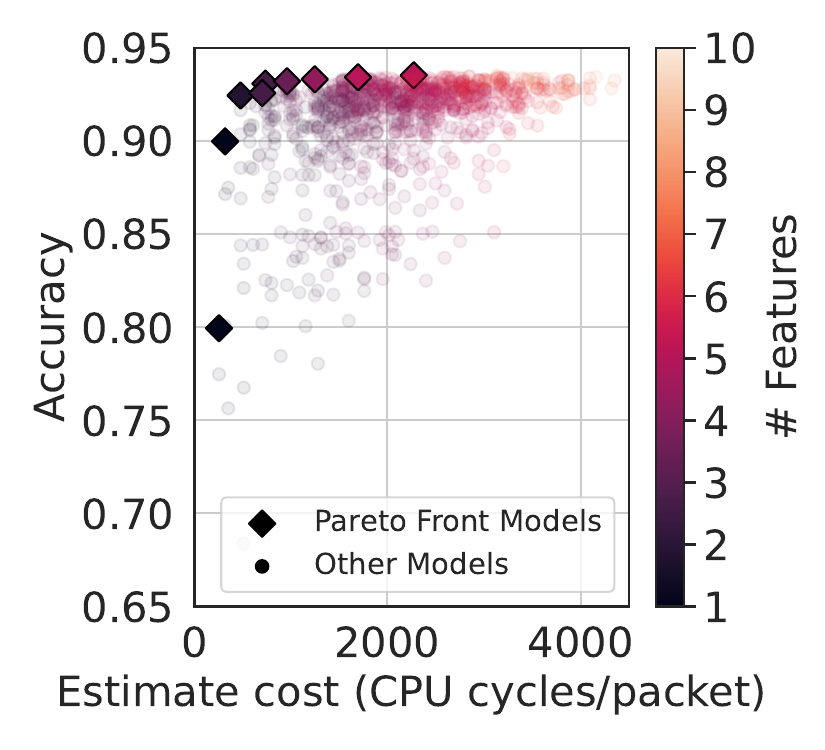}
\end{minipage}%

\begin{minipage}[b]{0.42\columnwidth}
    \captionof{table}{Video quality inference models.}
        \label{tab:featureset1}
\end{minipage}
\hfill
\begin{minipage}[b]{0.55\columnwidth}
    \captionof{figure}{Video quality inference Pareto front.}
        \label{fig:PF_video}
    \end{minipage}
\end{figure}

Table~\ref{tab:featureset1} shows the resulting configuration used as
system input. It consists of models labeled $m_X$, where $X$ represents the
accuracy-ordered index ($m_1$ being the least accurate and least computationally
expensive). To create this configuration, we applied CATO's methodology to
identify the Pareto front of candidate models illustrated in
Figure~\ref{fig:PF_video}. We started with 10 unitary features from
Bronzino~\etal~\cite{bronzino2019inferring}, representing data collected across
three protocol stack layers. For brevity, we omit the complete feature list for
each model, which includes common network metrics (packet inter-arrival time,
size, flow throughput) and video-specific features described in
Section~\ref{sec:context}. While the original paper contains more features, we
aggregated those computed from the same information (\eg, different statistical
representations of the same feature). Through CATO's approach, we reduced the
possible feature combinations from 1023 to just nine optimal configurations,
with those closer to the top left of the Pareto front being preferred.
Although \sysname{} leverages CATO's efficient computation method, it is
not tied to this specific algorithm—alternatives could be used, requiring
only that feature sets be ordered by accuracy and cost.

\subsection{Dynamic Feature Computation}\label{sec:feature_computation} 

The online phase of $\sysname$ consists of two key components, illustrated in
Figure~\ref{fig:diagram_CC}: the {\em Feature Computation} module and {\em Model
Selection} module. The Feature Computation module handles packet processing,
connection reassembly, feature extraction, and the computation of statistical
features required for ML model execution. The Model Selection module gathers
metrics to monitor system performance and determines the set of features to be
collected for the target model. We describe the Model Selection module in
Section~\ref{sec:model_selection}.

\paragraph{Packet processing parallelization.} 
Once incoming packets are received by the NIC, they are processed by the Feature
Computation module, which extracts the features required by the selected model,
aggregates them, and prepares them for consumption by the model. The Feature
Computation module consists of two primary components: Workers and a
Post-Processor. The Workers handle feature extraction directly from incoming
packets based on the feature set selected by the Model Selection module. Periodically,
the Post-Processor aggregates these features and exports them for use by the
model. Workers operate on dedicated CPU cores and perform feature extraction
tasks, including packet filtering, pre-processing (\eg, header parsing), and
basic feature computations (\eg, calculating flow throughput or packet
counters). \sysname{} leverages multi-core architectures to parallelize packet
processing across two dimensions. First, each Worker processes a subset of the
traffic. To account for features that depend on entire connections rather than
individual packets, all packets belonging to a given network flow are processed
by the same Worker. This is done via Receive Side Scaling (RSS).
Second, a backup Worker is instantiated to handle load while one of the Workers exports features to the Post-Processor.

The mapping between connections and Workers is managed through an indirection
table (illustrated in Figure~\ref{fig:diagram_CC}), which routes packets to the
appropriate Worker. Each Worker maintains a dedicated hashmap to store the
necessary data for monitored network flows. Each hashmap entry contains the set
of features computed for a specific connection, along with the required flow
state information. 

When a new packet enters the pipeline, it is directed to the corresponding
Worker. If the packet belongs to a new flow, the Worker first determines which
features need to be computed. It consults the shared configuration, which is
continuously updated by the Model Selection module (see
Figure~\ref{fig:diagram_CC}). This configuration dynamically specifies the
required features. Based on this information, the Worker initializes and
allocates the necessary data structures, which are then cached for future
packets from the same connection. For packets from known connections, the Worker
accesses the pre-existing structures associated with the connection, updates
relevant counters, extracts header information, and stores the data required for
the previously requested features. Since the selected features are determined
during the processing of the connection's first packet, the Worker does not need
to repeatedly access the configuration. This design choice ensures consistency
in feature computation for the duration of a network flow. However, this
approach comes with a trade-off: it limits the system's ability to quickly adapt
to changes in the selected features. Nevertheless, this trade-off is necessary
to maintain consistency, as certain features rely on historical data to be
computed (\eg, packet interarrival time distributions or video
segment size distributions).

\paragraph{Multi-task support.} Real-world scenarios typically require
simultaneous execution of multiple traffic analysis
tasks~\cite{vicenzi2023adaptiveinferenceonreconfigurablesmartnicsfortrafficclassification}.
Dedicating separate servers to each task is both inefficient and costly.
\sysname{} addresses this challenge by efficiently handling feature extraction
for multiple ML tasks concurrently while maintaining its key principles.
When a new connection appears, \sysname{} combines features set from
each task to avoid duplicate feature extraction and leverage intersections
between required features. This combination is enabled by designing required
features as composite sets of atomic features. Internally, each feature set
receives a unique identifier that can be decomposed into an N-bit
representation, with each bit corresponding to an atomic feature. During
processing, bits with value 1 trigger extraction of corresponding features,
while bits with value 0 are skipped. This approach allows straightforward
combination of feature sets from multiple tasks using a bitwise OR operation,
producing the final extraction set for incoming connections without additional
overhead or packet duplication. Additionally, since tasks may require different
packet quantities, the system extracts the higher number of packets requested.

\paragraph{Feature export.} 
Periodically, \sysname{} exports the computed features from the Workers to the
Post-Processor. The Post-Processor, itself running on a dedicated core, collects
the features extracted by the Workers, computes statistical features, and
formats the output for the machine learning model. The interval between two
exports is referred to as the $export\_window$. The act of exporting collected
features is often overlooked in existing work that emphasizes feature
extraction~\cite{CATO,ACDC}. However, this operation can be resource-intensive,
incurring significant CPU overhead and memory transfers, yet it is essential for
passing data to the ML model. 

To perform the export, incoming traffic processing on a Worker must be
temporarily halted to clear the hashmap storing the features. During this time,
Workers are unable to process incoming traffic, potentially leading to
significant packet drops if precautions are not taken. In our experiments, we
observed that this export process can take several seconds to complete. To
mitigate this, we exploit concurrent redundancy to implement a solution in
\sysname{}. When \sysname{} starts, a backup Worker is created, as shown in
Figure~\ref{fig:diagram_CC}. Then when $export\_window$ is reached, the monitor
swaps one of the current Workers and the backup Worker in the indirection
table. This will cause traffic to be redirected to the new Worker, then the
Monitor will notify the selected exporting Worker that it has been
swapped. It can then export its data to the Post-Processor without 
packet loss.

Once the export is complete, the Worker is designated as a backup, ready to
handle traffic during the next export cycle. At each $export\_window$, the
Worker clears its hashmap and sends the collected data to the Post-Processor.
The $export\_window$ interval depends on the specific use case and the hardware
capabilities, particularly the available memory. A shorter $export\_window$
requires less memory but introduces significant overhead. Conversely, a longer
$export\_window$ reduces overhead but risks memory saturation and may be
unsuitable if features must be delivered to ML models at shorter intervals.

\subsection{Adaptive Model Selection}\label{sec:model_selection}
The Model Selection module takes the pool of feature sets listed in the input
configuration and determines which model to use based on the ongoing load
experienced by the system. Many challenges lie behind this task. First,
monitoring the system's state cannot rely on heavyweight profiling that might
itself cause increased system load and lead to packet loss. Second, the system
must implement an intelligent algorithm capable of both detecting whether the
system is overloaded, thus triggering selection of a more lightweight feature
set, as well as whether the system has resources to spare, thus triggering a
switch to a more complex feature set. We solve the first challenge by relying on
a lightweight signal, \ie, packet loss, to monitor the system's state. This is
done by a dedicated Monitor that tracks the state of the \texttt{rx\_queues}
used to receive packets from the NIC. We monitor the
\texttt{rx\_miss} counter, which counts the number of packets dropped by the
hardware due to the queues being full.

\begin{algorithm}[t!]
    \small
    \caption{Feature set selection algorithm}
    \begin{algorithmic}[1]
        \State \textbf{mon\_window:} Time window for monitoring metrics
        \State $\mathrm{{\bf m_i:}}$ Index of the current candidate features set
        \State \textbf{n\_drops:} Dropped packets since the last cycle
        \State \textbf{dec\_factor:} Decrease factor used when a drop occurs 
        \State
        \Loop
        \If {$n\_drops > 0$}
        \State $m_i$ $\gets$ $\lfloor$ $m_i$ * $dec\_factor$ $\rfloor$
        \Else
        \If {$t\_since\_last\_update \geq mon\_window$}
        \State $m_i \gets m_i + 1$
        \State $t\_since\_last\_update \gets 0$
        \EndIf
        \EndIf
        \EndLoop
    \end{algorithmic}
    \label{algo:feature_selection}
\end{algorithm}

\paragraph{Methodology.}
We tackle the second challenge by designing an algorithm that leverages detected
losses—or their absence—to determine whether to switch between models. We
propose the algorithm outlined in Algorithm~\ref{algo:feature_selection} to
select a feature set based on the system's state. Note that \sysname{} is not
tied to this specific algorithm, and alternatives could be employed. The
candidate feature sets are indexed in increasing order of complexity, with the
currently selected set represented by the variable $m_i$. To prevent system
saturation, the algorithm continuously monitors for packet loss (line 7). When
loss is detected, $m_i$ is adjusted using a multiplicative decrease
($dec\_factor$), selecting a feature set with lower CPU cost. Conversely, the
algorithm additively increases $m_i$ to explore more complex feature sets, doing
so every $mon\_window$ seconds (line 10). This additive increase and
multiplicative decrease (AIMD) strategy is analogous to congestion control
mechanisms used in TCP algorithms that aggressively downgrade transmission rate
upon detecting a packet loss. Similarly, \sysname{} aggressively downgrades to a
simpler model when a loss is detected and cautiously explores more complex
models when no losses occur. The intuition behind this approach is that a
production deployment of \sysname{} could likely run multiple models in
parallel, each with different accuracy and CPU cost. As such, we seek
convergence to a balance of ideal feature sets that do not favor one model over
another. We evaluate our approach and compare it with alternatives in Section
\ref{sec:multitask}.

For \sysname{} to work effectively, the $mon\_window$ and $dec\_factor$
parameters must be carefully calibrated based on model accuracy requirements and
specific use cases. A higher $mon\_window$ value creates a more stable system by
causing more gradual transitions to complex feature sets. Conversely, a lower
$mon\_window$ enables quicker feature set upgrades but increases the risk of
system saturation and packet loss. The $dec\_factor$ determines how many feature
sets are skipped when drops are detected. A higher $dec\_factor$ reduces the
likelihood of subsequent losses by selecting significantly simpler feature sets,
though potentially sacrificing accuracy. After each change, the monitor updates
the Workers' feature sets as shown in Figure~\ref{fig:diagram_CC}. Finally, note
that for supporting multiple parallel tasks, the algorithm requires adaptation
to accommodate multiple $mon\_window$ values. This modification updates line 11
in Algorithm \ref{algo:feature_selection} by implementing a loop to test each
$mon\_window$ rather than just one. However, when drops are detected (line 7),
the model complexity ($m$) must be reduced for all tasks simultaneously.

\begin{figure}[t!]
    \includegraphics[width=\columnwidth,trim={8 5 8 0}, clip]{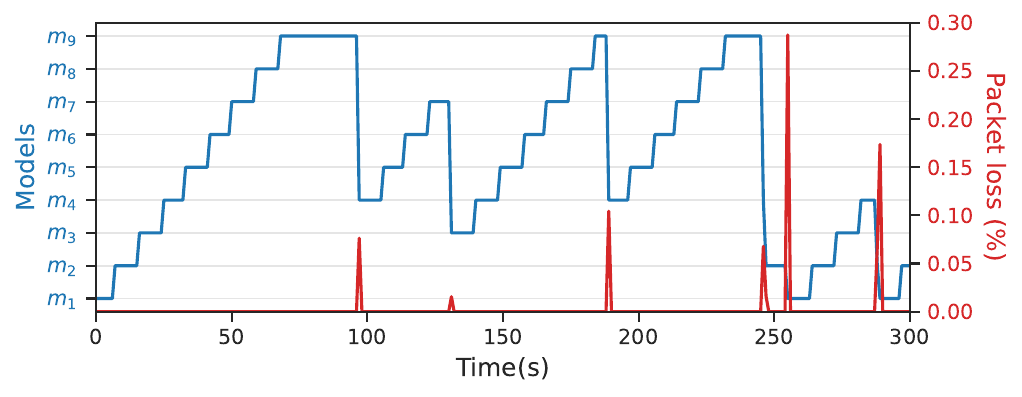}
    \caption{Timeseries of \sysname{} model selection algorithm. The blue line represents the selected feature set, while the red line shows the number of dropped packets over time.}
    \label{fig:TS_CC}
\end{figure}

\paragraph{Example.} To demonstrate the behavior of this algorithm in a
practical scenario, we run \sysname{} using a sample traffic trace: the first
five minutes of the CAIDA dataset previously utilized~\cite{caida2016}. The
results are presented in Figure~\ref{fig:TS_CC}. Each horizontal line in the
figure represents a feature set shown in Table~\ref{tab:featureset1}. The
accuracy of each corresponding model is shown on the left y-axis, with candidate
feature sets spanning accuracy values from 0.7994 ($m_1$) to 0.9353 ($m_9$). The
blue line illustrates the feature set selected by \sysname{} over time, while
the red line (right y-axis) represents the percent of dropped packets observed
every second.

As shown, we observe an initial phase where \sysname{} iteratively increases the
complexity of the computed feature sets, transitioning to more accurate models.
98 seconds into the experiment the system detects packet losses, prompting the
algorithm to rapidly downgrade to a less complex feature set. In this example,
with $dec\_factor$ set to 0.5, the algorithm switches from feature set $m_9$ to
$m_4$. When a second drop occurs at 132 seconds, the system further reduces
complexity, falling back to $m_3$. An interesting behavior appears at 247 seconds
where a transition to a less complex feature set occurs ($m_9$ to $m_4$), but the
system continues experiencing packet drops, immediately triggering a second
transition down to $m_2$. Finally, at 256 seconds, loss triggers a drop down to
$m_1$. It is important to note that these changes occur immediately upon
detection of packet drops by the monitor, rather than at each $mon\_window$
interval. The $mon\_window$ value is used exclusively for incrementing to more
complex feature sets.

\section{Prototype Implementation}\label{sec:eval} 
In this section, we present the proof-of-concept implementation of \sysname{}, as well as the two use case traffic analysis tasks currently implemented in the system release. 

\subsection{Software prototype}

We implement \sysname{} in Rust to leverage its features for ensuring memory and
thread safety while utilizing modern packet processing frameworks to maximize
packet processing efficiency. To handle high speed traffic, \sysname{} bypasses
the kernel to avoid bottlenecks caused by the network
stack~\cite{NetworkMeasurement,UnderstandingStackOverhead} using the Intel Data
Plane Development Kit (DPDK)~\cite{DPDK}. DPDK allows NIC to offload packets to
a dedicated CPU core's memory space without interruption by using Direct Memory
Access (DMA). Further, DPDK implements RSS to distribute the connection across
multiple CPU cores. We leverage this capability to share the connections between
multiple CPU cores and to swap traffic from one core to another when a worker
needs to export its hashmap to the post-processor.

Our prototype builds on a customized version of Retina\footnote{Retina v0.1 from
the original paper}, extending its monitoring core to support more complex tasks
such as dynamic worker reconfiguration and periodic feature export via updates
to the Indirection Table. We leverage Retina\cite{Retina} primarily for its
convenient and extensible Rust API for DPDK. However, we make minimal use of
Retina's higher-level features—relying mainly on basic packet filtering and the
DPDK runtime environment it provides. As a result, our system interacts closely
with core DPDK functionalities via Retina, which we believe could make it
straightforward to adapt to other DPDK-based environments. We implement a worker
that extracts monitor-specified features at line rate and exports them
periodically, and a post-processor that compute and formats these features for
machine learning models. The use of a dynamic trait ensures extensibility and
provides a shared interface that enables seamless coordination between the
worker, post-processor, and the feature set. In total, our prototype
implementation was \textasciitilde3300 lines of Rust code.

\subsection{Use Cases}\label{sec:use_cases}
We implement two typical traffic analysis tasks: video streaming quality
inference from encrypted traffic and service recognition.

\paragraph{Video quality inference.}\label{sec:video}
The first use case for \sysname{} focuses on inferring streaming video quality.
We implement the feature sets from Bronzino \etal~\cite{bronzino2019inferring},
prioritizing those identified by the Pareto front analysis in
Section~\ref{sec:offline}. These include basic network flow metrics (throughput
in/out, packet counts), end-to-end performance indicators (RTT, retransmission
statistics), and application-layer information derived from traffic patterns
(video segment sizes, inter-segment timing). Based on our ranked list, we
configure \sysname{} to operate with nine target models optimized for this use
case. This use case is particularly well-suited for our system, as it requires
continuous feature extraction throughout a flow's entire lifetime, making
performance heavily dependent on the total number of packets processed.

\paragraph{Service recognition.}\label{sec:detection}
For the second use case, we focus on service recognition, one of the most
studied traffic analysis tasks~\cite{BehindTheCurtain,AppIdentification}. This
task aims to determine the service (\eg, Netflix or YouTube) associated with a
given network flow. Accurate service recognition is critical for modern network
management, supporting performance monitoring (identifying flows associated with
a single application), quality of service (properly categorizing streaming
service flows), and security (detecting attack-related traffic). Widespread
encryption has made this process increasingly challenging, as packet payload
data becomes less informative after the initial TLS
handshake~\cite{shapira2021flowpic,liu2024serveflowfastslowmodelarchitecture}.
We focus on early flow detection, now a standard identification
approach~\cite{shapira2021flowpic,liu2024serveflowfastslowmodelarchitecture,babaria2025fastflow,piet2023ggfast},
utilizing just the first 10 packets of each connection. Unlike the video quality
inference use case, this task's collection cost is primarily determined by the
number of flows rather than packets, with the first packet carrying particular
significance due to memory structure initialization costs.

\begin{figure}[t!]
\begin{minipage}[c]{0.42\columnwidth}
        \centering
        \scriptsize
        \begin{tabular}{lrr}
            \toprule
            \textbf{Model \#} & \textbf{Cost} & \textbf{Accuracy} \\
            \midrule
            \rowcolor{Gray}$m_3$ & 1184 & 0.970 \\
            $m_2$ & 1056 & 0.898 \\
            \rowcolor{Gray}$m_1$ & 704 & 0.824 \\
            \bottomrule
        \end{tabular}
\end{minipage}
\hfill
\begin{minipage}[c]{0.55\columnwidth}
        \centering
        \vspace{4mm}
        \includegraphics[width=\columnwidth,trim={10 10 10 10},clip]{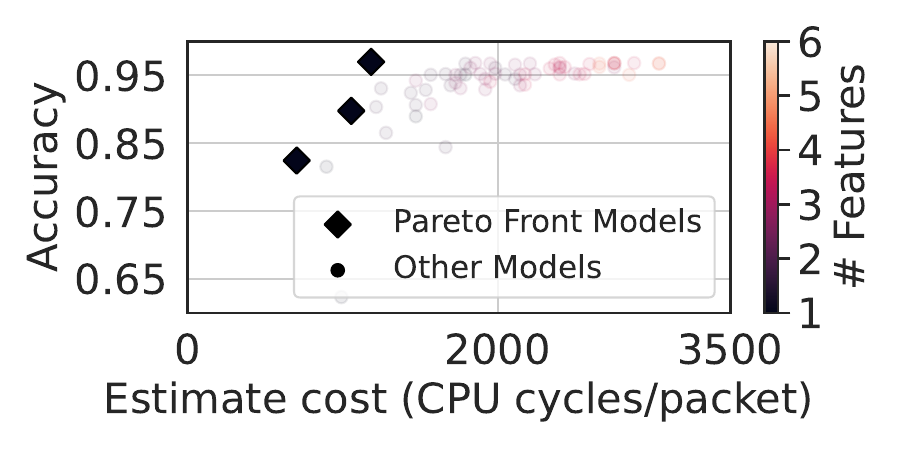}
\end{minipage}%

\begin{minipage}[b]{0.42\columnwidth}
    \captionof{table}{Service recognition models.}
        \label{tab:featureset2}
\end{minipage}
\hfill
\begin{minipage}[b]{0.55\columnwidth}
    \captionof{figure}{Service recognition Pareto front.}
        \label{fig:PF_recog}
    \end{minipage}
\end{figure}

We apply the same Pareto front methodology as in the previous use case. We
utilize six features from the video quality analysis, excluding the four video
segment identification features. Additionally, we consider raw headers (IP/TCP)
as a feature—a representation recently combined with deep learning models for
more accurate traffic identification~\cite{shapira2021flowpic,holland2021new}.
Raw headers require minimal computational complexity since no calculations are
needed for extraction, though they are more memory-intensive. This memory cost
is not reflected in our current CPU cycle-based metric\footnote{A cost metric
based on memory usage could also be used. We leave this for future
exploration.}. Consequently, raw headers offer high precision at low measured
cost, causing the Pareto front to be composed by only three target models. The
resulting Pareto front appears in Figure~\ref{fig:PF_recog} and
Table~\ref{tab:featureset2}, with three feature sets selected from the original
63. The raw headers feature set is represented as model $m_3$.

\section{Evaluation}\label{sec:eval} 
We evaluate \sysname's performance using several scenarios that allow us to
compare its performance in terms of accuracy and loss rates versus static
models, as well as to illustrate the effects of tunable parameters in
\sysname{}.
In all experiments, \sysname{} uses two Workers with only one active at a time, as presented in Section~\ref{sec:cruise_control}, except in the multicore experiments where multiple Workers run concurrently.
This setup simplifies the following analysis.

\paragraph{Hardware environment.} Our testbed consists of two identical servers,
each equipped with dual 16-core AMD EPYC 7343 processors and a 100GbE Intel E810
NIC with 384GiB of DDR4 split accros two NUMA.
Both servers are connected to a shared 100GbE switch. One server is
dedicated to traffic generation using Cisco's TRex~\cite{TRex}, which replays
real network traffic from capture files. This traffic is sent to the switch,
where it is mirrored to the second server running \sysname{}. The testbed
simulates a realistic environment in which a network operator monitors traffic
on a specific span port. It is worth noting that the use of AMD CPUs precludes
leveraging Intel's Data Direct I/O Technology (DDIO), which facilitates direct
transfers from the NIC to the CPU cache. 

\paragraph{Network traffic.} To mimic realistic traffic conditions, we use a
one-hour trace collected at Equinix Chicago in 2016~\cite{caida2016}. The
selected trace is a traffic capture available upon request from the CAIDA's
website (20160121). Throughout the trace, the throughput remains relatively
constant, with an average rate of approximately 10 Gbps. The trace's rate
fluctuates between 1 MPPS and 1.2 MPPS. To simulate different traffic rates in
our experiments, we adjust the inter-packet time using TRex. As with the
experiment discussed in Section~\ref{sec:context}, we use three different
traffic profiles: noon, evening, and night. The noon profile is used as the base
profile, with the evening profile scaled by a factor of 1.6 and the night
profile scaled by a factor of 0.2. Note that the trace contains intermittent
gaps of approximately one second. Although the exact origin of these gaps is
unknown, they may be caused by artifacts introduced during data capture or
anonymization.  They were retained in our dataset. This decision was made to
reflect realistic and potentially adverse conditions, under which ~\sysname{}
continues to perform robustly.

\paragraph{Evaluation objectives.} We design our experiments to answer the
following questions:

\vspace{1mm}\noindent\ding{182} {\em How does \sysname{} perform under varying
workloads?} We demonstrate that under realistic traffic conditions typical of
production deployments, \sysname{} improves median accuracy by 2.78\% while
reducing packet loss by a factor of four compared to statically-selected models,
the current state-of-the-art approach (\cref{sec:multipleworkload}).

\vspace{1mm}\noindent\ding{183} {\em What is \sysname{}'s overhead?} We show
that \sysname{} outperforms static configurations for both long-duration
experiments and short experiments that do not require feature exports
(\cref{sec:overhead}).

\vspace{1mm}\noindent\ding{184} {\em How scalable is \sysname{}?} We demonstrate
that \sysname{} effectively scales on multi-core architectures without incurring
excessive synchronization overhead (\cref{sec:multicore}).

\vspace{1mm}\noindent\ding{185} {\em Does \sysname{} ease the burden of parallel
feature collection?} We show that \sysname{} can be easily configured with
multiple parallel analysis tasks without significantly affecting model accuracy
or packet loss (\cref{sec:multitask}).

\vspace{1mm}\noindent\ding{186} {\em How do tunable parameters affect
\sysname{}'s performance?} We evaluate \sysname{}'s performance under different
$mon\_window$ values, selecting eight seconds as the optimal value for all other
experiments based on our findings (\cref{sec:monwindow}).

\subsection{Performance Under Varying Workloads}\label{sec:multipleworkload}
We evaluate \sysname{}'s ability to adapt to evolving traffic patterns typical
in real-world network deployments. Using three distinct traffic profiles (noon,
evening, and night) described earlier, we implement abrupt transitions between
profiles to assess the system's responsiveness to sudden load changes. We
compare \sysname{} against state-of-the-art systems like Retina~\cite{Retina}
and CATO~\cite{CATO}, which require pre-selecting features before deployment and
need substantial downtime to reconfigure. For our experiments, we use a modified
version of Retina with feature export capabilities—a necessary adaptation to
prevent memory exhaustion during extended tests (detailed in
Section~\ref{sec:overhead}). We test Retina using all optimal configurations
identified from our Pareto front analysis and label results according to the
model used. We perform experiments using a single Worker core, as our goal is to
compare \sysname's performance against static configurations. We evaluate
multi-core scalability in Section~\ref{sec:multicore}.

\begin{figure}[t!] 
    \centering
    \begin{subfigure}[t]{0.5\columnwidth} 
        \includegraphics[width=\textwidth]{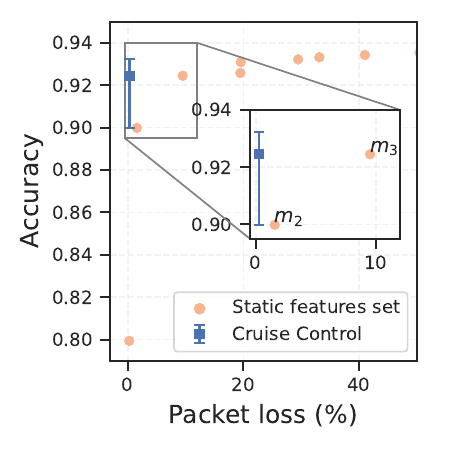}
        \vspace{-0.2in}
        \caption{\centering Video quality inference} 
        \label{fig:IQR_video_modified_speed}
    \end{subfigure}%
    \begin{subfigure}[t]{0.5\columnwidth} 
        \includegraphics[width=\textwidth]{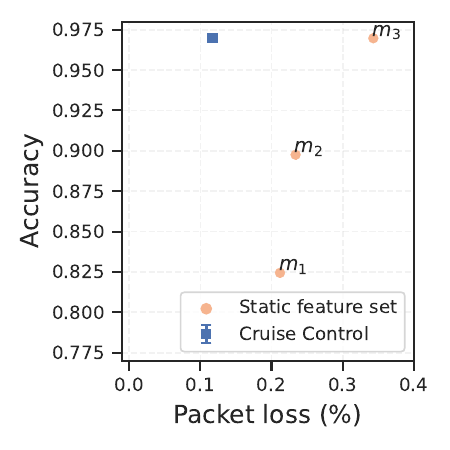}
        \vspace{-0.2in}
        \caption{Service Recognition\vspace{1.2em}}
        \label{fig:IQR_recog_modified_speed}
    \end{subfigure} 
    \hfill
    \vspace{-0.1in}
    \caption{Different time of day workload}
\end{figure}

\paragraph{\sysname{} improves both accuracy and packet loss.}
Figure~\ref{fig:IQR_video_modified_speed} illustrates the performance
comparison, plotting accuracy (y-axis) against packet loss rates (x-axis).
Orange circles represent static configurations (Retina), while the blue square
denotes \sysname{}. Note that, since CAIDA traffic is unlabeled, we cannot
directly evaluate machine learning model performance—which is beyond the scope
of this paper. Instead, ``accuracy'' represents the relationship between
extracted features and expected model performance as determined in the offline
phase (Section \ref{sec:cruise_control}). For \sysname{}, which dynamically
employs multiple feature sets, we show median accuracy with first and third
quartiles. The optimal configuration would appear in the top-left corner,
indicating maximum accuracy with zero packet packet loss. 

For the video quality inference use case
(Figure~\ref{fig:IQR_video_modified_speed}), all static feature sets beyond $m_2$
incur excessive packet loss rates (>9\%), making them impractical for
deployment. In contrast, \sysname{} achieves just 0.37\% loss while matching
$m_3$'s median accuracy. Even $m_2$, which might be selected in static deployments
to avoid unacceptable losses, performs worse than \sysname{} with higher loss
(1.57\%) and lower accuracy (0.025 difference in median). For the service
recognition use case (Figure~\ref{fig:IQR_recog_modified_speed}), all
configurations experience lower packet losses compared to the first use case,
attributable to the simpler analysis requirements (processing only the first 10
packets per connection). Nevertheless, \sysname{} still outperforms static
configurations in accuracy and packet loss reduction.

\begin{figure*}[t!] 
    \centering 
    \begin{subfigure}[b]{.49\textwidth}
        \centering 
        \includegraphics[width=\textwidth]{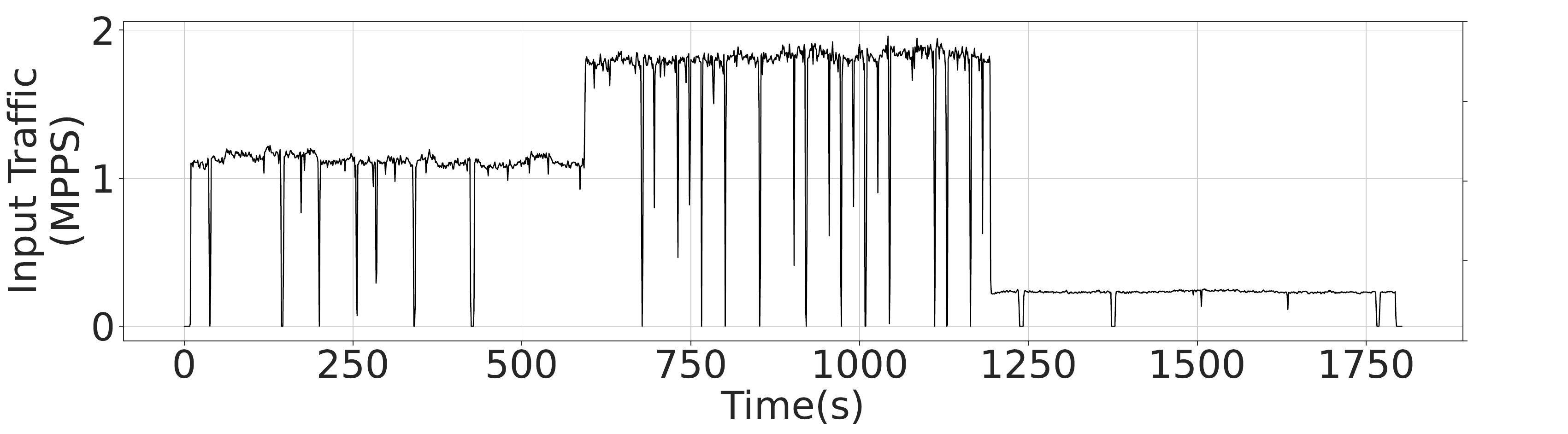} 
        \caption{Injected traffic} 
        \label{fig:injected}
    \end{subfigure}\hfill
    \begin{subfigure}[b]{.49\textwidth} 
        \includegraphics[width=\textwidth]{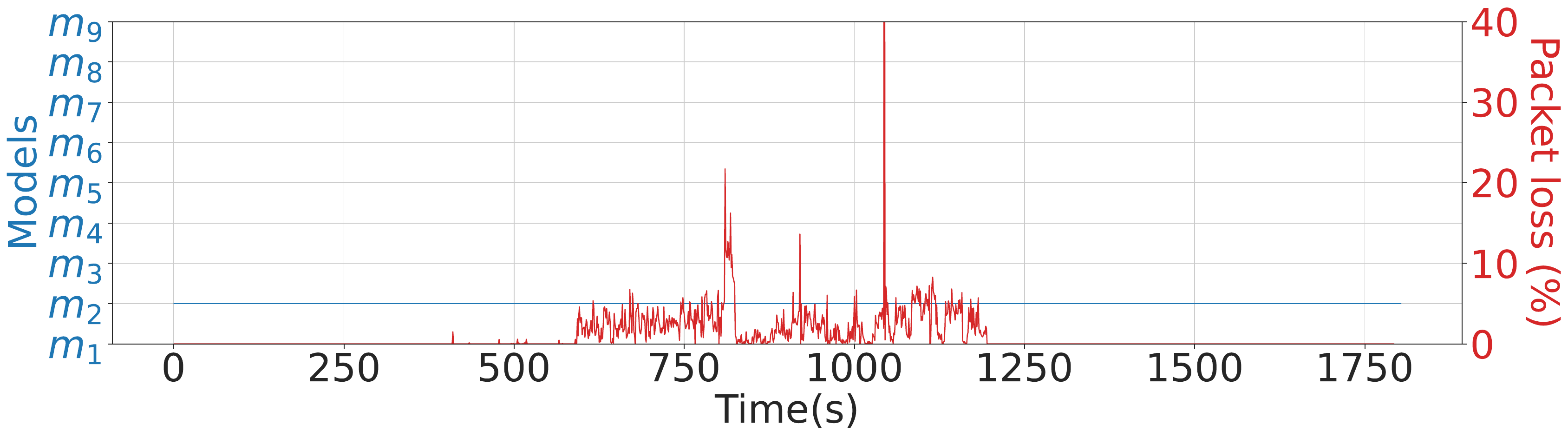} 
        \caption{$m_2$}
        \label{fig:m2}
    \end{subfigure} \hfill
    \begin{subfigure}[b]{.49\textwidth} 
        \centering
        \includegraphics[width=\textwidth]{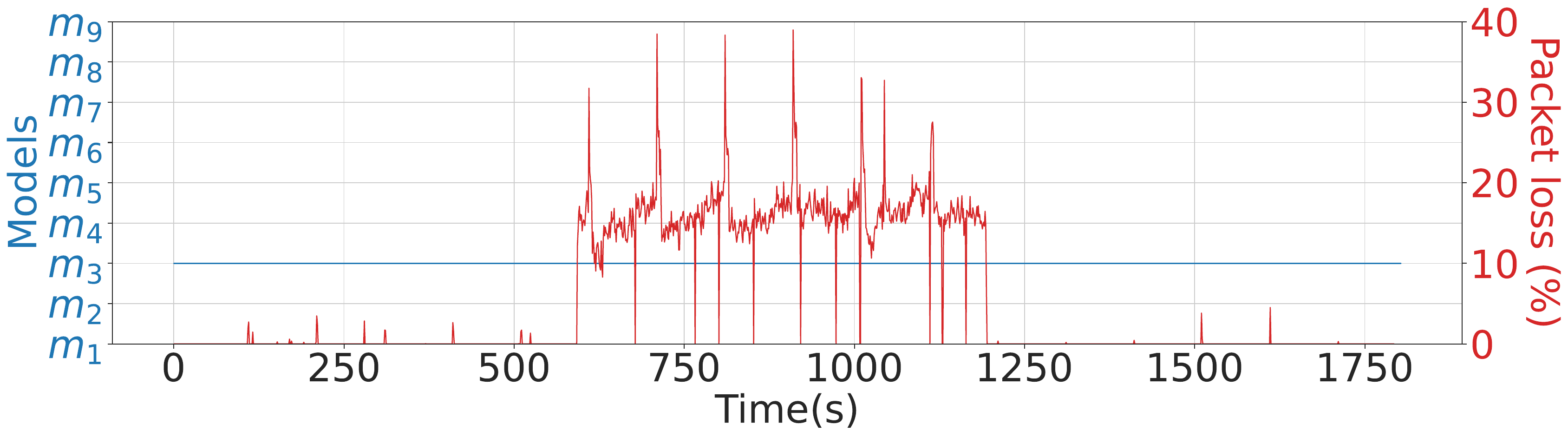}
        \caption{$m_3$}
        \label{fig:m3}
    \end{subfigure} \hfill
    \begin{subfigure}[b]{.49\textwidth} 
        \centering
        \includegraphics[width=\textwidth]{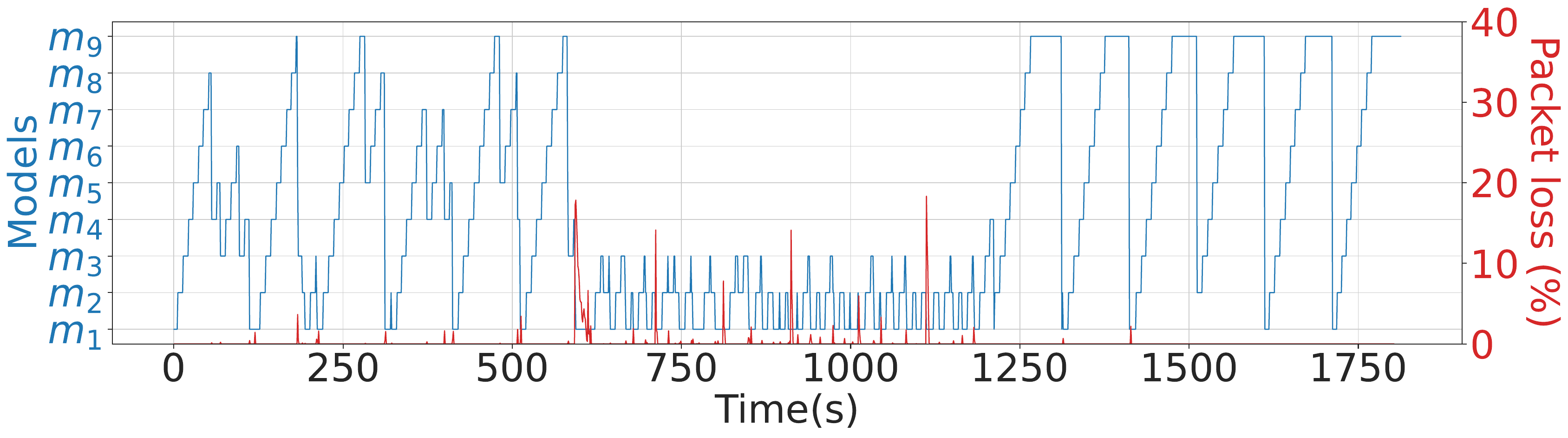}
        \caption{\sysname{}}
        \label{fig:cc}
    \end{subfigure} 
    \vspace{-0.1in}
    \caption{Timeseries for static feature sets and $\sysname$ for video quality inference}
    \label{fig:TS_video}
\end{figure*}

\paragraph{\sysname{} self-adapts to sudden workload changes to prevent packet loss.}
We further investigate these results by examining the causes of packet loss,
using the first use case as a representative example. Figure~\ref{fig:TS_video}
presents a time series of the traffic profile (Figure~\ref{fig:injected})
alongside system behavior data for both \sysname{} (Figure~\ref{fig:cc}) and two
representative static configurations (Figures~\ref{fig:m2} and \ref{fig:m3}).
These graphs display model accuracy (blue/left y-axis) and packet losses
(red/right y-axis) throughout the experiment. With the $m_3$ static feature set,
minimal drops occur during the 'noon' traffic profile (0-600 seconds), primarily
coinciding with export events. During the 'evening' profile (600-1200 seconds),
the system cannot compute $m_3$ for all packets, resulting in significant losses.
The 'night' profile (final 10 minutes) processes without loss. We plot the
average accuracies of $m_2$ and $m_3$ as horizontal blue lines at 0.899 and
0.9245, respectively. However, static configuration accuracy becomes
unpredictable during packet loss due to randomness, particularly evident during
peak traffic (600-1200 seconds) where $m_2$ experiences approximately 5\% loss
and $m_3$ reaches 15\%.

In contrast, \sysname{} adapts dynamically, as shown in Figure~\ref{fig:cc}.
During the first 10 minutes, it progresses through the Pareto front models to
reach the most accurate ones, stepping down to less complex models when losses
occur. At the 600-second mark, when traffic increases, a burst of packet loss
triggers \sysname{} to adjust to the least accurate (but least costly) model
$m_1$, which successfully processes traffic without loss, thus performing even better than $m_2$.
When traffic decreases at 1200 seconds, \sysname{} reverts to the most accurate model. This experiment
demonstrates \sysname{}'s ability to adapt to changing loads while minimizing
losses compared to static configurations. Though \sysname{} introduces some
accuracy variability, it effectively prevents and responds to packet loss while
achieving higher accuracy during periods of excess processing capacity.

\begin{table}[t!]
    \centering
    \footnotesize
    \begin{tabular}{lrrrrrr}
        \toprule
         \textbf{Dataset} &  & $m_3$ & $m_4$ & $m_5$ & $m_6$&\textbf{CC}\\
        \midrule
        \rowcolor{Gray}\textbf{CAIDA} &
        \textbf{Pkt loss (\%)}  & 0.047 & 0.357 & 0.490  & 13.041 & \textbf{0.033} \\
        & \textbf{Accuracy}& 0.924 & 0.926 & 0.931 & 0.932 & \textbf{0.931} \\
        \midrule
        \rowcolor{Gray}\textbf{No Export} &
        \textbf{Pkt loss (\%)} & 0 & 0.064 & 0.062  & 11.393 & \textbf{0.043} \\
        & \textbf{Acc} & 0.924 & 0.926 & 0.931 & 0.932  & \textbf{0.931} \\
        \bottomrule
    \end{tabular}
    \caption{Performance comparison between static model and ~\sysname{} (CC) for video quality inference}
    \label{tab:UC1}
\end{table}

\begin{table}[t!]
    \centering
    \footnotesize
    \begin{tabular}{lrrrrr}
        \toprule
        \textbf{Dataset} & &  $m_1$ & $m_2$ & $m_3$& \textbf{CC} \\
        \midrule
        \rowcolor{Gray}  \textbf{CAIDA} &
       \textbf{Pkt loss (\%)}  & 0.212 & 0.234 & 0.343 & \textbf{0.117}\\
        & \textbf{Accuracy} & 0.824  & 0.898  & 0.970 & \textbf{0.970} \\
        \midrule
     \rowcolor{Gray}  \textbf{No Export} &
        \textbf{Pkt loss (\%)} &  0 & 0.001 & 0.013 & \textbf{0}\\
        & \textbf{Accuracy} & 0.824  & 0.898  & 0.970 & \textbf{0.970} \\
        \bottomrule
    \end{tabular}
    \caption{Performance comparison between static model and ~\sysname{} (CC) for service recognition}
    \label{tab:UC2}
\end{table}

\begin{table*}[t!]
    \centering
    \footnotesize
    \begin{tabular}{lrrrrrrrrr}
        \toprule
        \textbf{\# Parallel Workers} & \textbf{ 0.5 Mcps } & \textbf{ 1.0 Mcps } & \textbf{ 1.5 Mcps } & \textbf{ 2.0 Mcps } & \textbf{ 2.5 Mcps } & \textbf{ 3.0 Mcps } & \textbf{ 4.0 Mcps } & \textbf{ 6.0 Mcps } & \textbf{ 8.0 Mcps } \\
        \midrule
        \rowcolor{Gray} 1 &  0.00& \textbf{0.00}& 0.79& 23.87& 30.39& 40.13& 55.73& 70.48& 77.62\\
        2 &  0.00& 0.00& 0.00& 0.00& \textbf{0.00}& 2.90& 23.66& 56.50& 67.24\\
        \rowcolor{Gray} 4 &  0.00& 0.00& 0.00& 0.00& \textbf{0.00}& 5.64& 17.50& 26.74& \textit{29.45}*\\
        8 &  0.00& 0.00& 0.00& 0.00& 0.00& \textbf{0.00}& 1.21& \textit{0.00}*& \textit{5.88}*\\
        \bottomrule
    \end{tabular}
    \caption{Median packet loss (\%) evaluation on multi-core during 10 minutes experiment. *\textit{Memory exhausted}}
    \label{tab:multicore}
\end{table*}

\subsection{System Overhead}\label{sec:overhead}

Our previous experiment demonstrates \sysname{}'s advantages under rapidly changing traffic conditions. Here, we show that \sysname{} also outperforms static configurations for both longer, steadier workloads and shorter workloads that do not require feature exports.

\paragraph{Long workloads.}
We evaluate our system using a complete 1-hour CAIDA trace with a steady traffic
rate of approximately 1.1 MPPS ($4.2 \times 10^9$ packets total).
Table~\ref{tab:UC1}, section CAIDA, shows that \sysname{} performs well in this
realistic scenario. Overall, it experiences minimal packet loss—lower than the
static feature set $m_3$ (by about 0.05\%)—while primarily utilizing $m_5$, which
loses 0.49\% of packets. For the second use case (Table~\ref{tab:UC2}, section
CAIDA), \sysname{} primarily uses $m_3$ while providing almost $3\times$ lower
packet loss. As noted earlier, some minimal losses during exports are
inevitable—the necessary trade-off for reducing end-to-end latency and enabling
longer experiments without memory exhaustion. However, this experiment
demonstrates that \sysname{} effectively handles longer, steadier workloads
while maintaining low packet loss rates and high accuracy.

\paragraph{Short workloads.}
To verify that our results are not artifacts created by \sysname's export
mechanism, we conduct a scaled-down experiment limited by available RAM. This
test includes only five minutes of traffic with all features kept in memory,
eliminating any overheads related to hashmap transfers to the post-processor.
Results appear in Tables~\ref{tab:UC1} and~\ref{tab:UC2} (No Export section).
\sysname{}'s behavior remains consistent with previous experiments. In this
scenario, \sysname{} experiences slightly more packet loss (0.043\%) than $m_3$
(0\%), but performs better than $m_4$ and $m_5$ (0.062\% and 0.064\% loss,
respectively) while achieving median accuracy equal to $m_5$. This again
highlights \sysname{}'s advantage over static configurations—significant
accuracy improvements with minimal packet loss penalties. For the second use
case (Table~\ref{tab:UC2}), we observe insignificant packet loss across (less than 0.001) 
all models except $m_3$, which experiences drops at startup due to processing 
only the first 10 packets of each flow. 

\subsection{Multi-core Scalability}\label{sec:multicore}
Our previous experiments focused on single-core scenarios to demonstrate
\sysname{}'s adaptability to changing workloads. We now assess system
scalability through a dedicated multi-worker (\ie, multi-core) benchmark. We
focus on the worst-case scenario: new connection arrivals. In this test, each
packet represents a new connection, forcing the system to perform high-cost
operations including configuration checks, data structure creation, and storage
in thread-local hash maps.  The following results are presented in Millions of connections per
second (Mcps).

Table \ref{tab:multicore} shows that the system scales effectively with
increasing worker counts. Moving from one to two workers, throughput doubles
from 1 to 2 million connections per second (Mcps), indicating near-perfect
scaling. At four workers, throughput reaches 2.5 Mcps, indicating diminishing
returns relative to ideal linear scaling. This behavior could potentially be
attributed to unfairness of RSS\cite{Barbette2019RSS++}, we leave exploration of
this to future work. Notably, with eight workers, the system reaches 3 Mcps,
significantly exceeding real-world traffic levels, such as those observed in
CAIDA traces, which report only 23.92 thousand new connections per second. As
these tests represent worst-case scenarios where each packet establishes a new
connection, we anticipate superior performance in realistic environments.

It is worth noting that experiments at 6 Mcps and 8 Mcps on eight workers, and 8
Mcps on four workers, terminated prematurely due to memory exhaustion before the
allocated 10-minute duration. This highlights the importance of careful memory
management when utilizing multiple workers. The primary challenge involves
thread-local hashmaps—while providing processing advantages through
worker-specific storage, improperly dimensioned hashmaps require runtime
resizing, resulting in packet processing failures and consequent losses.
Additionally, in the experiments referenced in Table \ref{tab:multicore}, we
apply identical parameters across different worker counts. Since only one
worker exports at a time (determined by round-robin rotation), each of the N
workers exports after a total duration of $N \times export\_window$.
Consequently, with more workers, each remains active longer before exporting,
necessitating larger hashmaps and increasing the load on the
worker-to-post-processor channel. Neglecting this consideration leads to rapid
memory exhaustion, while excessive switching increases overhead costs.

\begin{figure*}[tbp] 
    \centering 
    \begin{subfigure}[b]{0.49\textwidth} 
        \includegraphics[width=\textwidth]{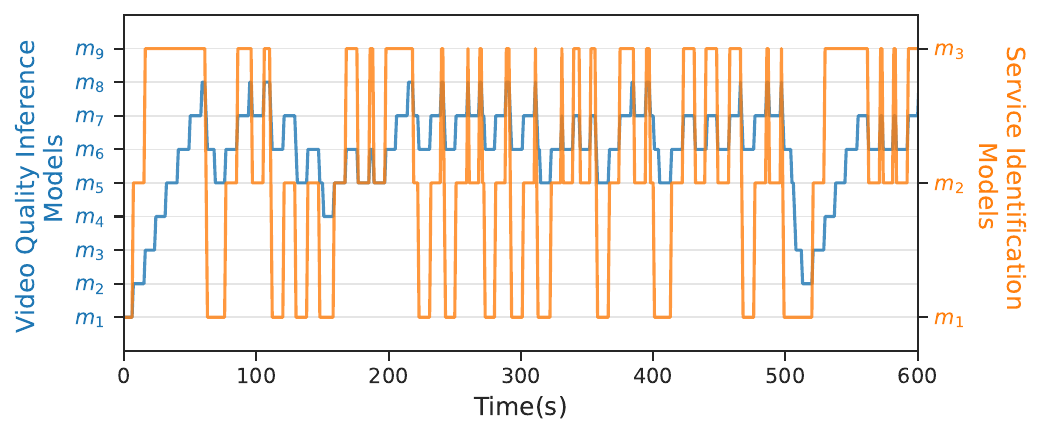} 
        \vspace{-0.25in}
        \caption{(AIAD,AIAD)} 
        \label{fig:AIAD_AIAD}
    \end{subfigure}%
    \hspace{0.1in}
    \begin{subfigure}[b]{0.49\textwidth} 
        \includegraphics[width=\textwidth]{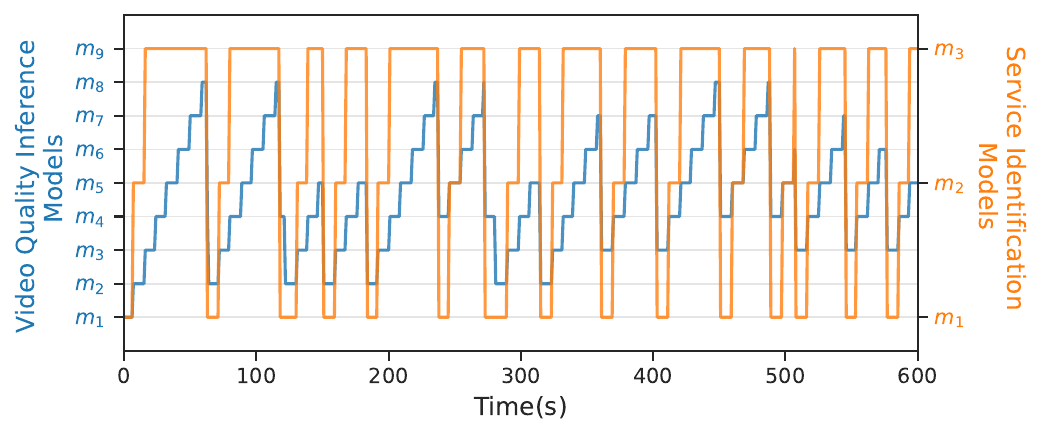} 
        \vspace{-0.25in}
        \caption{(AIMD,AIMD)} 
        \label{fig:AIMD_AIMD}
    \end{subfigure}
    \vspace{-0.2in}
    \caption{Service recognition features extraction across three different network load}
    \label{fig:multitask}
\end{figure*}

\subsection{Multitask Support}\label{sec:multitask}
While extracting features for a single ML task demonstrates our system's basic
capabilities, \sysname{} is designed to support feature extraction for multiple
concurrent tasks. In this experiment, we demonstrate this capability by running
both use cases in parallel on the first 10 minutes of the previously used CAIDA
trace. In the following, $UC1$ stands for video quality inference and $UC2$ for service recognition.
Our objectives are twofold: to show that the system maintains high model
accuracy and low packet loss while supporting multiple tasks, and to establish
that AIMD is the most appropriate
control algorithm for this scenario. We accomplish this by conducting four
experiments comparing AIMD against an Additive Increase Additive Decrease (AIAD)
approach that decrement by one $m_i$ when drop occurs, unlike AIMD’s multiplicative decrease.
Note that, although the algorithm supports task-specific $mon\_window$ values as described in
Section~\ref{sec:cruise_control}, we apply the same value to both tasks based on
our finding that this configuration delivers optimal performance for both use
cases (detailed in the following section).

Table~\ref{tab:multitask} shows a summary of the experiment for the four
combinations. We observe that using AIAD tends to increase overall accuracy but
also results in a higher packet drop rate. Both outcomes can be explained by
AIAD’s slower reaction to packet drops, causing longer periods of packet loss
but smaller decreases in overall performance. The (AIMD, AIAD) combination
achieves higher accuracy with only a limited packet drop penalty. This is
probably because $UC2$'s Pareto front includes only three models, making each
step more impactful. As a result, AIAD appears sufficient. Nevertheless, the
(AIMD, AIMD) combination demonstrates the most favorable results in minimizing
packet drops.

To better understand the reasons behind these results, we plot in Figure
\ref{fig:AIAD_AIAD} and \ref{fig:AIMD_AIMD} the two opposite scenarios, \ie,
using AIAD in both versus using AIMD in both. The blue line represent the
selected configuration for $UC1$ while the red line the configuration selected
for the $UC2$. We add AIAD to the AIMD to test multiple combinations. 
We observe that (AIAD, AIAD) results in weaker synchronization between the two tasks, with $UC1$ leveraging its broader set of models on the Pareto front to benefit from this imbalance. 
In contrast, (AIMD, AIMD) shows a more regular sawtooth pattern, where both tasks synchronize more effectively.

Overall, these results confirm the choice of AIMD as the more effective control
algorithm for exploration to limit the packet loss, even complex scenarios such
as multitasking.

\begin{table}[t!]
    \centering
    \footnotesize
    \begin{tabular}{rr|rrrr}
        \toprule
        \textbf{UC1} & \textbf{UC2} & \textbf{Acc UC1} & \textbf{Acc UC2} & \textbf{Total Drop (\%)}\\
        \toprule
        \rowcolor{Gray}AIAD & AIAD & 0.932 & 0.824 & 0.0263 \\ 
        & AIMD & 0.932 & 0.824 & 0.0251 \\   
        \rowcolor{Gray}AIMD & AIAD & 0.931 & 0.970 & 0.0089 \\
        & AIMD & \textbf{0.926} & \textbf{0.970} & \textbf{0.0082} \\ 
        \bottomrule 
    \end{tabular}
    \caption{Median Accuracy and Total Drop(\%) of different combinations of control algorithms}
    \label{tab:multitask}
\end{table}

\subsection{Sensitivity to Parameters}\label{sec:monwindow} 
Finally, we evaluate how the $mon\_window$ parameter impacts system performance.
This key parameter controls how quickly \sysname{} switches to more complex
configurations when no packet loss is detected, serving as a performance tuning
lever that operators can adjust based on specific deployment goals (\eg,
maximizing accuracy or minimizing packet loss). We conduct ten experiments using
the 1-hour CAIDA traffic trace. Table~\ref{tab:UC1_mon_window} summarizes the
packet loss percentages and median accuracy results across various $mon\_window$
values. Both metrics decrease with increasing $mon\_window$ as the system remains
longer on each model/feature set before switching. For this particular use case
and trace, an eight-second $mon\_window$ provides the optimal balance between
accuracy and packet loss. Similar experiments with our second use case show
consistent behavior—both accuracy and packet loss decrease with larger
$mon\_window$ values. The optimal value remains eight seconds (yielding 0.147\%
packet loss and 0.97 accuracy). We recommend conducting comparable experiments
for any use case to determine the optimal $mon\_window$ value according to
specific operator requirements.

\begin{table}[t!]
    \centering
    \footnotesize
    \begin{tabular}{lrrrrrrrrrr}
        \toprule
        \textbf{$mon\_window$} & \textbf{1} & \textbf{2} & \textbf{4} & \textbf{8} & \textbf{10} \\
        \midrule
        \textbf{Packet loss (\%)} & 0.107 & 0.074 & 0.048 & \textbf{0.033} & 0.031  \\
        \textbf{Accuracy} & 0.932 & 0.931  & 0.931  & \textbf{0.931} & 0.926 \\
        \bottomrule
    \end{tabular}
    \caption{Impact of $mon\_window$ on ~\sysname{}}
    \label{tab:UC1_mon_window}
\end{table}

\balance
\section{Related work}
Many studies have investigated similar individual subcomponents
integral to \sysname{}. Here we examine the related work and discuss key
differences with \sysname{}'s design.

\paragraph{System for extracting features from network traffic.}
Network feature extraction has been a dynamic research area for decades, with
systems designed to derive meaningful insights from traffic. Recent advancements
focus on high-speed network traffic processing. For example,
PacketMill~\cite{PacketMill} optimizes software for 100Gb/s throughput,
Enso~\cite{Enso} introduces a streaming abstraction for improved efficiency, and
Retina~\cite{Retina} uses multilayered filtering and streamlined feature
extraction for relevant network flows. However, these solutions rely on static
configurations that can lead to suboptimal performance under changing network
conditions. In contrast, \sysname{} introduces a dynamic configuration framework
that adapts to varying network loads in real time, offering a more flexible
approach.

\paragraph{Cost-aware ML model creation.}
Neglecting system constraints during model training can significantly impact
inference time, causing packet loss that affects ML model performance. To
address this, Traffic Refinery~\cite{TrafficRefinery} proposes a methodology to
explore and mitigate data representation technical debt, but it requires manual
intervention. In contrast, CATO~\cite{CATO} performs automated, end-to-end
optimization of the traffic analysis pipeline, but it's offline and requires
model selection online. Liu \etal~\cite{liu2024serveflowfastslowmodelarchitecture} underscore the impact of
inter-packet time on the feature extraction process, proposing an approach using
three parallel models with varying packet requirements. However, they solely
focus ib early application identification and using static models. Other work
explores in-network inference within programmable devices such as
switches\cite{parizotto2023SurveyPISA,xiong2019SwitchMl,akem2024encryptedtrafficclassificationatlinerateinprogrammableswitcheswithmachinelearning}
or
FPGAs~\cite{vicenzi2023adaptiveinferenceonreconfigurablesmartnicsfortrafficclassification,Elnawawy2020FPGA}
to exploit high-speed processing capabilities. However, these approaches face
limitations due to restricted command and extraction capabilities of in-network
fabric, constraining model features and sophistication.

\paragraph{Dynamic model selection.}
Few studies explore dynamic model selection for traffic analysis.
pForest~\cite{bussegrawitz2022pforest} dynamically switches between multiple ML
models based on flow packet count, while Jiang \etal~\cite{ACDC} investigate
selecting from classifiers with varying feature requirements to balance
classification speed and memory consumption, focusing primarily on memory usage
costs. In contrast, \sysname{} emphasizes CPU cycles and tailors model selection
to optimize computational performance. 
Vicenzi et al.~\cite{vicenzi2023adaptiveinferenceonreconfigurablesmartnicsfortrafficclassification} propose an adaptive framework that switches between pruned convolutional neural network (CNN) variants based on accuracy and throughput.
They use a fixed set of features. 
Our approach, on the other hand, adapts the feature set itself.
Additionally, their system focus on FPGAs, while ours targets commodity hardware.
Beyond traffic analysis, adaptive model selection has been
explored in general ML serving systems. Clipper~\cite{crankshaw2017clipper} and
INFaaS~\cite{romero2021INFaas} dynamically optimize model selection based on
performance requirements. Zhang \etal~\cite{zhang2020modelswitch} pioneered
dynamic model serving based on load and pre-characterized performance, inspiring
recent
studies\cite{jeon2025SLO,Salmani2023reconcilinghighaccuracycost-efficiencyandlowlatencyofinferenceservingsystems}
on machine learning as a service. These works focus on avoiding
service-level objective violations with financial penalties, whereas \sysname{}
focuses on system-level computational costs for processing
features for network traffic analysis.

\section{Conclusion}\label{sec:conclusion} In this work, we introduce
\sysname{}, a system that dynamically selects target ML models for traffic
analysis tasks at runtime, without requiring user intervention. \sysname{}
leverages lightweight signals to adapt to changing network conditions and the
system's available resources. We detail the design of \sysname{} and evaluate it
across two use cases, demonstrating its advantages under varying traffic
configurations. 

\sysname{} opens several promising directions for future research.
One avenue is improving model cost estimation by incorporating additional performance and resource metrics.
Another is identifying new system-level metrics to enable finer-grained model selection and improve performance.
We believe the presented approach has the potential to inspire
new research directions, advancing the deployment of machine learning-based
tasks in operational networks. To encourage further exploration and innovation,
we are releasing \sysname{} as an open-source project.

\label{lastpage}
\newpage
\microtypesetup{protrusion=false}

\bibliography{paper.bib}


\vfill
\pagebreak

\end{document}